\documentstyle[preprint,aps,pra]{revtex}
\tightenlines
\begin{document}



\draft

\title{
First and second sound in a uniform Bose gas}

\author{A. Griffin}
\address{Department of Physics, University of Toronto\\ Toronto,
Ontario, Canada M5S 1A7} 

\author{and}
\address{}

\author{E. Zaremba}
\address{Department of Physics, Queen's University \\ Kingston,
Ontario, Canada K7L 3N6}

\date{\today}

\maketitle

\begin{abstract}
We have recently derived
two-fluid hydrodynamic equations for
a trapped weakly-interacting Bose gas.  In this paper, we use these
equations to discuss first and second sound in a uniform Bose gas.
These results are shown to agree with the predictions of the usual
two-fluid equations of Landau when the thermodynamic functions
are evaluated for a weakly-interacting gas. In a uniform gas,
second sound mainly corresponds to an oscillation of the superfluid
(the condensate) and is the low frequency continuation of the
Goldstone-Bogoliubov symmetry-breaking mode.
\end{abstract}


\section{Introduction}\label{sect1}
One of the most spectacular features \protect\cite{ref1} exhibited
by
superfluid $^4$He is the existence of two hydrodynamic sound modes,
first and second sound.  As first pointed out by Tisza
\protect\cite{tis40},
the motion of a Bose condensate as a separate degree of freedom
results in a two fluid hydrodynamics describing the superfluid and
normal fluid components\protect\cite{lan41}.  In recent work, the
authors \protect\cite{zargrinik}
gave a microscopic derivation of the two-fluid hydrodynamic
equations of motion for a trapped weakly-interacting Bose-condensed
gas.  In contrast to a Bose-condensed liquid like superfluid
$^4$He, the superfluid in a gas corresponds directly to the
condensate atoms and the
normal fluid corresponds to the non-condensate (or excited) atoms. 
In the present paper, we use these two-fluid equations to discuss
the first and second sound modes of a {\it uniform} Bose-condensed
gas.  We find that at temperatures close to $T_{BEC}$,  first
(second) sound mainly corresponds to an
oscillation of the non-condensate (condensate) atoms.  We also
confirm \protect\cite{gaygri85,gri93} that it is the second sound
mode in a
uniform gas which is the low frequency hydrodynamic analogue of the
collisionless Bogoliubov-Popov Goldstone mode \protect\cite{pov65}.

These results for a uniform gas are of interest for comparison with
the hydrodynamic  
oscillations of the condensate and non-condensate in a {\it
non-uniform}
trapped Bose gas \protect\cite{zargrinik,zargri}. 
They also may be of direct interest in connection with recent
studies  at MIT  \protect\cite{andetal} of  the propagation of
pulses along the
z-axis of a cigar-shaped trap. The axial trap spring constant is so
small that the condensate along the z-axis can be treated as
effectively uniform (to a first approximation) in such propagation
studies.

We recall that  Ref.~\protect\cite{zargrinik} (ZGN) is based on:
(a) a
time-dependent 
Hartree-Fock-Popov equation of motion for the condensate
wavefunction $\Phi({\bf
r},t)$; and (b) a set of hydrodynamic equations for the
fluctuations of the thermal cloud (non-condensate) based on a
kinetic equation which includes the effect on the atoms of the 
time-dependent self-consistent Hartree-Fock field.
The analysis of ZGN uses the local equilibrium solution of
the kinetic equation and thus
does not include any hydrodynamic damping, such as Kirkpatrick and
Dorfman \protect\cite{kirdor} consider. 
However it should be emphasized that a local equilibrium
description is crucially dependent on collisions between the atoms
and thus the hydrodynamic equations are only valid for low
frequency phenomena ($\omega\ll 1/\tau_c$, where $\tau_c$ is the
mean time between collisions of atoms in the thermal cloud).

In  Section \ref{sect2}, we solve the linearized hydrodynamic
 two-fluid equations for the coupled superfluid and normal fluid
velocity fluctuations derived in ZGN. We exhibit the first and
second sound normal modes valid at intermediate temperatures,
defined as the temperature regime below $T_{BEC}$ where the
interaction energy of an atom is much less than the thermal
kinetic energy (i.e., $gn_0 \ll k_BT$; here, $n_0$ is the gas
density and $g=4\pi a\hbar^2/m$ is the interaction parameter). The
analysis of ZGN is built on a mean-field approximation for the
equilibrium properties.  As discussed in Section \ref{sect3}, this
simple theory is not valid close to the superfluid transition, where
it gives rise to spurious discontinuities in the condensate
density.  In Section \ref{sect4}, we discuss the relation between
our two-fluid equations written in terms of velocity fluctuations
and the standard Landau formulation given in terms of density and
entropy fluctuations \protect\cite{ref1,lan41}.  All previous
discussions \protect\cite{leeyan59,pov65,gaygri85} of 
hydrodynamic modes in a dilute
Bose gas have used the latter formulation.

\section{Coupled equations for superfluid and normal fluid
velocities}\label{sect2}

When there is no trapping potential,
the non-condensate density $\tilde n_{0}$ and condensate density
$n_{c0}$ do
not depend on position.  In this case, one can reduce the
linearized 2-fluid equations given by Eqs.~(12), (15) and (16) of
ZGN to two coupled equations for the normal
and superfluid local velocities
\begin{mathletters}
\label{eq1}
\begin{eqnarray}
m{\partial ^2\delta{\bf v}_S\over\partial
t^2}&=&gn_{c0}\bbox{\nabla}
(\bbox{\nabla}\cdot\delta{\bf v}_S) +
2g\tilde n_{0}\bbox{\nabla}(\bbox{\nabla}\cdot\delta{\bf
v}_N)\label{eq1a}\\
m{\partial ^2\delta{\bf v}_N\over\partial t^2}&=& 
\left ( {5\over 3} {\tilde P_0\over\tilde n_{0}} + 2g\tilde n_0
\right ) \bbox{\nabla} (\bbox{\nabla}\cdot\delta{\bf v}_N) +
2gn_{c0}\bbox{\nabla}(\bbox{\nabla}\cdot\delta{\bf v}_S)\ .
\label{eq1b}
\end{eqnarray}
\end{mathletters}

\noindent 
We emphasize that these equations are only valid at finite
temperatures such that $gn_0 \ll k_BT$.
In deriving these equations, we have assumed that the contribution 
from the first
term of Eq.(13) of ZGN is negligible in the long-wavelength limit
of interest. These equations can be solved
to give the low frequency hydrodynamic normal modes of a uniform
Bose-condensed gas,
 as will be discussed.  We defer discussion of  the equilibrium
quantities ($n_{c0}$, $\tilde n_0$ and 
the kinetic contribution to the pressure $\tilde P_0$) which appear
in (\ref{eq1a})
and (\ref{eq1b}) to Section \ref{sect3}.

Introducing the velocity potentials $\delta {\bf v}_S\equiv
\bbox{\nabla} \phi_S$,
$\delta{\bf v}_N \equiv \bbox{\nabla} \phi_N$, it is easy to see
that (\ref{eq1a}) and
(\ref{eq1b}) have plane-wave solutions $\phi_{S,N}({\bf r},t) =
\phi_{S,N} e^{i({\bf k}\cdot {\bf r} -\omega t)}$ satisfying 

\begin{eqnarray}
\left [\omega^2 - {gn_{c0}\over m}k^2\right ] \phi_S -
\left({2g\tilde n_{0}\over m }k^2\right) \phi_N &=&0\nonumber\\
&&\nonumber\\
-\left ({2gn_{c0} \over m}k^2\right )\phi_S + \left [ \omega^2 -
\left
({5\over 3} {\tilde P_0 \over m\tilde n_{0}} + {2g\tilde n_{0}
\over
m}\right )k^2\right ] \phi_N &=& 0\ .
\label{eq6}
\end{eqnarray}

\noindent The zeros of the secular determinant of this coupled set
of equations give two phonon solutions $\omega_{\pm}^2 = 
u_{\pm} ^2 k^2$, where the velocities are the solution of 
\begin{equation}
u^4 - u^2 \left ( {5\over 3} {\tilde P_0 \over m\tilde n_0} +
{2g\tilde n_0 \over m} + {gn_{c0} \over m} \right )
+ {gn_{c0} \over m} \left ( {5\over 3} {\tilde P_0 \over m\tilde
n_0} 
- {2g\tilde n_0 \over m} \right )= 0 \, .
\label{eq6'}
\end{equation}
Expanding to second order in the explicit dependence on $g$, the
sound velocities 
are given by
\begin{mathletters}
\label{eq7}
\begin{eqnarray}
u^2_+ &=& {5\over 3} {\tilde P_0\over m\tilde n_{0}} + {2g\tilde
n_{0}\over m} +
{gn_{c0} \over m} \epsilon
\label{eq7a}\\
u^2_- &=&{gn_{c0}\over m} - {gn_{c0}\over m} \epsilon\ ,
\label{eq7b}
\end{eqnarray}
\end{mathletters}

\noindent where $ \epsilon\equiv 4 g\tilde n_{0}/{ \textstyle
{5}\over
\textstyle{ 3}}{\textstyle {\tilde P_0 }\over
\textstyle {\tilde n_{0}}} \ll 1$ is the expansion parameter.  
We note (see Section III) that the ratio ${\textstyle{\tilde P_0}\over
\textstyle{\tilde n_{0}}} = k_BT {\textstyle {g_{5/2}(z_0)}\over
\textstyle{g_{3/2}(z_0)}}$ depends weakly on $g$. 
 The $\omega_+$ mode in (\ref{eq7a})  clearly corresponds to first
sound.  Using $\omega_+^2 = u^2_+ k^2$ in (\ref{eq6}), one finds
to leading order in $g$ that
\begin{equation}
{\phi_N\over \phi_S}\simeq {2\over \epsilon}  \gg 1 \ .
\label{eq8}
\end{equation}

\noindent That is to say, the $\omega _+$ first sound mode
corresponds
to an {\it in-phase} oscillation in which the non-condensate
velocity
amplitude is much larger than that of the condensate.
The $\omega_-$ mode in (\ref{eq7b})  is the second sound mode. 
Using $\omega_-^2 =
u_-^2 k^2$ in (\ref{eq6}), one finds to leading order in $g$ that
\begin{equation}
-{\phi_S\over\phi_N}\simeq {2\over \epsilon} {\tilde n_0 \over
 n_{c0}} \gg 1 \ .
\label{eq9}
\end{equation}

\noindent Thus at finite temperatures where (\ref{eq1a}) and
(\ref{eq1b}) are valid, second sound in a uniform
weakly-interacting gas is
seen to be an {\it out-of-phase} oscillation, in which the
condensate
velocity amplitude is much larger than that of the non-condensate
(a similar result was obtained many years ago in
Ref.~\protect\cite{leeyan59}).

\section{Equilibrium properties in the Popov
approximation}\label{sect3}

The ZGN derivation of the coupled hydrodynamic equations for the
two velocity fields given in (\ref{eq1a}) and (\ref{eq1b}) is built
on a self-consistent Hartree-Fock description of the equilibrium
properties.  One of the earliest discussions of this mean-field
theory was given by Popov \protect\cite{pov65} and it has become the
standard approximation in recent studies of trapped Bose gases. 
Referring to ZGN, we recall that the equilibrium equation for the
condensate yields the
equilibrium chemical potential
\begin{equation}
\mu_0=2g\tilde n_{0} +gn_{c0}\,.
\label{eq2}
\end{equation}
This parameter enters in the determination of the
equilibrium excited-atom density given by
\begin{equation}
\tilde n_{0}(T,n_0) = {1\over\Lambda^3} g_{3/2}(z_0)\,,
\label{eq3}
\end{equation}
where ($n_0\equiv n_{c0}+\tilde n_0$)
\begin{equation}
z_0 = e^{\beta(\mu_0-2gn_{0})} = e^{-\beta gn_{c0}}
\label{eq4}
\end{equation}
is the equilibrium fugacity and $\Lambda = 
\sqrt{2\pi \hbar^2/mk_B T}$ is the thermal de Broglie wavelength.
The associated excited-atom {\it kinetic} pressure is
\begin{equation}
\tilde P_0(T,n_0) = {1\over\beta\Lambda^3} g_{5/2}
(z_0)\ .
\label{eq5}
\end{equation}

\noindent We note that these results are equivalent to the simple
``toy model'' studied in Ref.\protect\cite{huayanlut57}.

Eqs.(\ref{eq3}) and (\ref{eq4}) must be solved self-consistently
to determine $n_{c0}$ and $\tilde n_0$ for a given total density
$n_0$. Condensation occurs when the density
reaches the critical density $n_{cr} = g_{3/2}(1)/\Lambda^3$.
For $ n_0 < n_{cr}$, the condensate density is zero and (\ref{eq3})
with $\tilde n_0 = n_0$ determines the
equilibrium fugacity. In Fig.\ 1, we show the equilibrium
densities as a function of volume for a fixed temperature. 
The parameter $\gamma_{cr} \equiv \beta g n_{cr}$ is used to
characterize the strength of the interaction.
We see that the present level of approximation leads to a
discontinuous change in the densities at the transition 
point\cite{foot1}.
Moreover, below the critical volume $v_{cr} = 1/n_{cr}$, $\tilde
n_0$
decreases as a result of the interactions with the condensate,
in contrast to the ideal gas behaviour which has the
non-condensate maintaining a constant density of $n_{cr}$.
Fig. 2 gives the total pressure defined as \protect\cite{zargrinik}
\begin{equation}
P = \tilde P_0 + {1\over 2} g (n_0^2 + 2n_0 \tilde n_0 -\tilde
n_0^2)\ ,
\label{eq5'}
\end{equation}
normalized by the critical pressure $\tilde P_{cr} =
g_{5/2}(1)/\beta\Lambda^3$ of the ideal gas.
The second term in (\ref{eq5'}) is the {\it explicit} interaction
contribution, but it
should be noted that $\tilde P_0$ also depends on
interactions as a result of its dependence on $z_0$. The
discontinuous behaviour of the non-condensate density leads to an
analogous discontinuity in the pressure\cite{foot1}. In Figs. 3
and 4 we show the corresponding behaviour as a function of $T$.
It is of interest to note that for a trapped Bose
gas, the use of these equilibrium properties
in the Thomas-Fermi approximation leads to a
similar discontinuous behaviour of the equilibrium condensate 
density, but now as a function of the radial distance from the 
center of the trap\cite{goldman}.

It is clear that the properties
of the weakly-interacting gas are nonanalytic functions of the 
interaction strength $g$ at the transition point within the
mean-field Popov approximation described by
 (\ref{eq2})--({\ref{eq5}).
However, one should not take these features in the BEC critical
region 
seriously.  The simple mean-field Popov approximation for
interactions  is well known \protect\cite{pop83,shi97} not to
be valid very close to the transition and the predicted
discontinuities
exhibited in Figs. 1--4 (characteristic of a first-order
transition) are indicative of the limitations of the 
present simple theory.
A correct treatment of this region would require a renormalization
group (RG) analysis\protect\cite{bylsto96}
which is outside the scope of the present paper.

For later purposes, we note that the kinetic pressure $\tilde
P_0$ in (\ref{eq5}) can be calculated by expanding the fugacity 
as $ z_0 \simeq 1 - \beta g n_{c0} + \cdots$, which yields
(using the identity $z\partial g_n(z)/\partial z = g_{n-1}(z)$)
\begin{equation}
\tilde P_0 \simeq \tilde P_{cr} - g n_{c0} n_{cr}\ ,
\label{press}
\end{equation}
where $\tilde P_{cr}$ and $n_{cr}$ are the critical pressure and
density of the ideal Bose gas introduced earlier.  However, a
similar
perturbative
expansion of the non-condensate density $\tilde n_0$ in (\ref{eq3})
is not
possible since the derivative of $g_{3/2}(z)$ diverges at $z
=1$. Indeed, it is this non-perturbative dependence on $g$ which
leads to the discontinuities shown in Figs.~1--4.

\section{Relation to standard two-fluid equations}\label{sect4}
First and second sound in a uniform Bose-condensed gas have been
previously discussed in the 
literature\protect\cite{gaygri85,pov65,leeyan59}. 
These earlier treatments start
with the usual two-fluid equations of Landau \protect\cite{lan41}. 
We recall that these linearized equations are (see ch.7 of
Ref.\protect\cite{ref1})
\begin{eqnarray}
{\partial \delta n\over \partial t} &=& -\bbox{\nabla}\cdot \delta
{\bf j}\nonumber\\
m {\partial \delta {\bf v}_S\over \partial t} &=&
-\bbox{\nabla}\delta\mu\nonumber\\
m {\partial\delta {\bf j}\over \partial t} &=& -\bbox
{\nabla}\delta P\nonumber\\
{\partial \delta s\over \partial t} &=& - \bbox{\nabla} 
\cdot (s_0\delta {\bf v}_N)\ ,
\label{neweq11}
\end{eqnarray}

\noindent where 

\begin{eqnarray}
\delta n({\bf r}, t)& = &\delta \tilde n({\bf r}, t) + \delta
n_c({\bf r}, t)\nonumber\\
\delta {\bf j} ({\bf r}, t) &=& \tilde n_0\delta {\bf v}_N + 
n_{c0}\delta {\bf v}_S\ .
\label{neweq12}
\end{eqnarray}
$P$ and $s$ are the pressure and entropy density, respectively. 
ZGN proved that the two-fluid equations which lead to
(\ref{eq1a}) and (\ref{eq1b}) are in fact {\it equivalent} to the
two-fluid
equations in (\ref{neweq11}) when the thermodynamic functions in
the latter are evaluated for the present model of a
weakly-interacting Bose gas. Using the thermodynamic relation 
\protect\cite{zargrinik}, $n_0\delta\mu =\delta P-s_0\delta
T$, to eliminate the chemical potential, and defining the
entropy per unit mass by $\bar s \equiv s/m n = s/\rho$,
one can reduce the equations in (\ref{neweq11}) to
\protect\cite{ref1,lan41}

\begin{eqnarray}
m {\partial ^2\delta n\over \partial t^2}&=&\nabla ^2\delta
P\nonumber\\
{\partial^2\delta \bar s\over \partial t^2} 
&=&{\rho_S\over \rho_N} \bar s_0^2\nabla^2\delta T\ .
\label{eq13}
\end{eqnarray}

\noindent Solving this closed set of equations in terms of the
variables $\delta n$ and $\delta \bar s$, one finds two normal mode
solutions $\omega^2 \equiv u^2k^2$, where $u^2$ is given by the
solution of the quadratic equation \protect\cite{lan41}
\begin{equation}
u^4-u^2\left [ \left . {\partial P\over \partial \rho
}\right\vert_T
+ {T\over \bar c_v} \left (
{1\over \rho} \left . {\partial P\over \partial T}\right\vert_\rho
\right )^2 
+ {\rho_S\over \rho_N}
{T\bar s_0^2\over \bar c_v}
\right ] +
{\rho_S\over\rho_N} {T\bar s_0^2\over \bar c_v}
\left . {\partial P\over \partial \rho}\right\vert_T =0\,.
\label{eq14}
\end{equation}
In this equation, $\bar c_v$ is the specific heat per unit mass
and derivatives of the pressure have been expressed in terms of
the independent thermodynamic variables $T$ and $\rho$. Although
not immediately apparent, 
the coefficients in (\ref{eq14}) are in
fact consistent with those appearing in (\ref{eq6'}). 

The problem is thus reduced to evaluation of the various
equilibrium thermodynamic functions and derivatives which appear in
(\ref{eq14}).  
For the entropy per unit mass we have the 
expression\protect\cite{zargrinik}
\begin{equation}
\rho_0 \bar s_0 T = {5\over 2}\tilde P_0 + g\tilde n_0
n_{c0}\ ,
\end{equation}
from which we obtain
\begin{equation}
\rho_0 \bar c_v = {3 \over 2} \rho_0 \bar s_0
+ g \left ({3\over 2} \tilde n_0 + n_{c0} \right ) \left .
{\partial \tilde n \over \partial T}\right\vert_\rho \,.
\end{equation}

\noindent From the equation of state (\ref{eq5'}), we find that
\begin{equation}
\left . {\partial P \over \partial \rho}\right\vert_T = 
{gn_0 \over m} \left (
1+ \left . {\partial \tilde n \over \partial n}\right\vert_T \right
)
\end{equation}
and 
\begin{equation}
\left . {\partial P \over \partial T}\right\vert_\rho = 
\rho_0 \bar s_0
+ g n_0 \left . {\partial \tilde n \over \partial
T}\right\vert_\rho \,.
\end{equation}
These quantities have been calculated previously in the limit
that the interaction parameter $g$ is regarded as small
\protect\cite{gaygri85,leeyan,grekle}. 
In this situation, $\tilde P_0$ in (\ref{eq5}) is approximated by
(\ref{press}).  An additional approximation is typically
made whereby $\tilde n_0$ is simply replaced  by the ideal gas
expression $n_{cr}$, in which case
$n_{c0} = n_0 - n_{cr}$. To the same level of approximation, one
finds
$\left . {\partial \tilde n \over \partial n} \right\vert_T = 0$
and $\left . {\partial \tilde n \over \partial T} \right\vert_\rho
= 3n_{cr}/2T$. With these replacements, we also note that the
expressions
for the pressure and the entropy and energy densities given in
ZGN reduce precisely to those of Refs. \protect\cite{leeyan}
and \protect\cite{grekle}. 

Using these results to calculate the thermodynamic quantities in
(\ref{eq14}), the
first and second sound velocities are found (after some algebra)
to be given by

\begin{mathletters}
\label{eq11}
\begin{eqnarray}
u^2_+& =&{5\over 3} {k_BT\over m} {g_{5/2}(1)\over g_{3/2}(1)} +
{2g n_{cr}\over m} - {5\over 3} {gn_{c0}\over m}
\label{eq11a}\\
&&\nonumber\\
u^2_- &=& {gn_{c0}\over m}\ ,
\label{eq11b}
\end{eqnarray}
\end{mathletters}

\noindent keeping terms to first order in $g$.  The leading order
terms in (\ref{eq11a}) and (\ref{eq11b}) were  obtained from
(\ref{eq14}) by this
method by Popov (see the last paragraph of Ref.
\protect\cite{pov65}) as well as
by Lee and Yang \protect\cite{leeyan59}. Precisely the same results
follow from (\ref{eq7}) to first order in $g$ when (\ref{press})
is again used for the kinetic pressure $\tilde P_0$ and $\tilde
n_0$ is replaced by $n_{cr}$. However, the
results given by (\ref{eq6'}) are more general than those in 
(\ref{eq11}), which only keep the leading order corrections
to the properties of a  non-interacting gas.  As we discussed
above, the 
analysis leading to (\ref{eq11}) 
ignores any interaction-correction to the non-condensate density
$\tilde n$, which as can be seen from Fig.\ 1, becomes
significant as the density increases beyond $n_{cr}$.

As we emphasized in the beginning of Section \ref{sect2}, the
analysis of ZGN assumes that $gn_0 \ll k_BT$ and thus the results
are not really valid at low temperatures.  To  discuss the low
temperature region would require a generalization of our work which
is based on a quasiparticle spectrum exhibiting phonon-like
behavior at long wavelengths (such a kinetic equation has been
derived in Ref.\protect\cite{kirdor}). The pioneering work of Lee
and Yang \protect\cite{leeyan59} did include an analysis of both
the low temperature and high temperature regions.  At low
temperatures, they found that the first and second sound modes
avoid becoming degenerate by hybridizing and an interchange of
the physical meaning of these two modes occurs as a result of this
hybridization. While the sound velocities
given by (\ref{eq6'}) are not really valid at low
temperatures, Fig.\ 5 shows that our results do lead to this expected
hybridization of first and second sound in a dilute gas.

\section{Concluding remarks}\label{sect5}

Recently two-fluid hydrodynamic equations were derived
\protect\cite{zargrinik} for a trapped, weakly-interacting Bose
gas. These are given in terms of coupled equations for the
superfluid and normal fluid velocity fluctuations.  In order to
obtain more physical insight into these hydrodynamic equations,
we have given in the present paper a detailed analysis for
a {\it uniform} Bose gas.  In this case, it has
been proven that the hydrodynamic equations of
Ref.\protect\cite{zargrinik} are formally equivalent to the usual
Landau two-fluid equations. As the present paper shows,
this formal equivalence is somewhat hidden in explicit
calculations of the first and second sound velocities. However, as
discussed in Section \ref{sect4}, our results do reduce (to first order
in the interaction $g$) to those found in earlier studies
\protect\cite{leeyan59,pov65,gaygri85} based on the Landau
formulation.

In superfluid $^4$He, one evaluates the equilibrium thermodynamic
parameters in (\ref{eq14}) using the phonon-roton excitation
spectrum.
  As is well known\protect\cite{ref1,lan41}, in
superfluid $^4$He, first sound corresponds to an in-phase
oscillation in
which ${\bf v}_N ={\bf v}_S$.  In contrast,
second sound corresponds to an out-of-phase oscillation in which
$\rho_n{\bf
v}_N = -\rho_S{\bf v}_S$.  The
difference between second sound in a dilute gas at
finite temperatures and in a liquid  is a result of the dominance
of the
kinetic energy over the interaction energy for atoms in a gas.  In
both cases, however, we note that  the second sound frequency goes
to zero (becomes soft) at the
superfluid transition.  The mode does {\it not} exist above $T_{BEC}$. 
Moreover (\ref{eq7b}) shows that second sound crucially depends on
the
interaction $g$.  It would be absent if we had set $g=0$ in
(\ref{eq1a}) and (\ref{eq1b}).

As we have noted, second sound in a dilute gas largely involves an
oscillation of the condensate atoms (superfluid density) and is a
soft mode which vanishes in the normal phase.  We recall that at
finite temperatures \protect\cite{pov65}, the generalization of the
$T=0$ Bogoliubov phonon gives a velocity formally identical to the
first
term in (\ref{eq7b}).  Thus we conclude that in a weakly
interacting Bose-condensed gas at finite temperatures, second sound
is the low frequency
(hydrodynamic regime) continuation of the high frequency
(collisionless or mean-field regime) Bogoliubov-Goldstone mode.
This was first suggested in Refs.
\protect\cite{gaygri85,gri93,pov65}. 
The situation is quite different in superfluid $^4$He, where the
collisionless phonon spectrum is the continuation of hydrodynamic
first sound \protect\cite{gri93} and there is no high-frequency
analogue of the second sound branch.

\acknowledgements
We thank W. Ketterle for a copy of Ref.\ \protect\cite{andetal}
before publication and useful remarks.
This work was supported by research grants from NSERC of Canada.


\vfil\break
\centerline{\bf FIGURE CAPTIONS}
\begin{itemize}
\item[Fig.1:]
Density vs. volume per particle for a fixed temperature $T$. 
The chain curve corresponds to
the non-condensate, the solid curve to the condensate.
$\gamma_{cr}$ is the value of $gn_0/k_B T$ at the critical
density $n_{cr} = g_{3/2}(1)/\Lambda^3$.

\item[Fig.2:]
Pressure isotherms: the solid line is the total pressure
according to (\ref{eq5'}), the
chain curve is $\tilde P_0$ and the dashed curve corresponds to
the usual approximation\cite{leeyan,grekle}
$P \simeq \tilde P_{cr} + {1\over 2} g(n^2 + n_{cr}^2)$.

\item[Fig.3:]
As in Fig. 1, but as a function of $T$ for a fixed density
$n_0$. Here, $\gamma_{cr}=gn_0/k_B T_{BEC}$.

\item[Fig.4:]
Normalized pressure as a function of $T$ for a fixed density
$n_0$. The solid curve corresponds to (\ref{eq5'}) and the chain
curve is $\tilde P_0$. The dashed curve below $T = T_{BEC}$ is the
ideal gas result $\tilde P_0/\tilde P_{cr} = (T/T_{BEC})^{5/2}$.

\item[Fig.5:]
Squares of the first and second sound velocities (normalized by
the first sound velocity of the ideal gas at $T = T_{BEC}$) 
vs. $T/T_{BEC}$. The value of $\gamma_{cr}$ has been increased
to more clearly reveal the anti-crossing behavior at low
temperatures. As discussed in Section IV, the low temperature
results only indicate the qualitative behavior.
\end{itemize}

\begin{references}
\bibitem{foot1} Within the present model, the discontinuity in the
density at the transition 
is $\Delta \tilde n/n_{cr} = 4\pi \gamma_{cr} / 
[g_{3/2}(1)]^2$, to lowest order in the parameter 
$\gamma_{cr} \equiv g n_{cr}/k_B T$. 
Similarly, the discontinuity in the pressure is
$\Delta P/\tilde P_{cr} = 4\pi \gamma_{cr}^2/g_{3/2}(1)g_{5/2}(1)$.
\end{references}
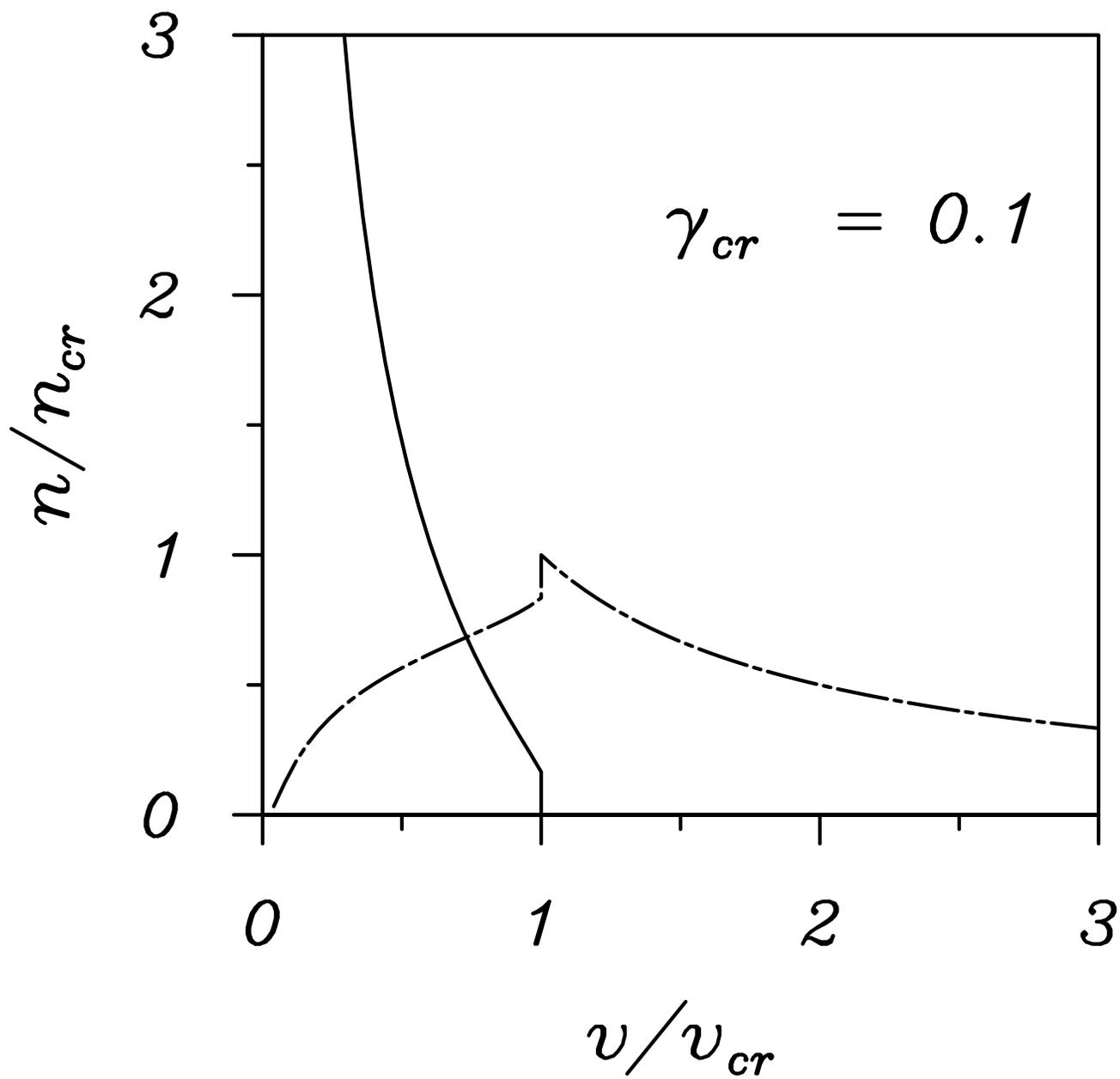
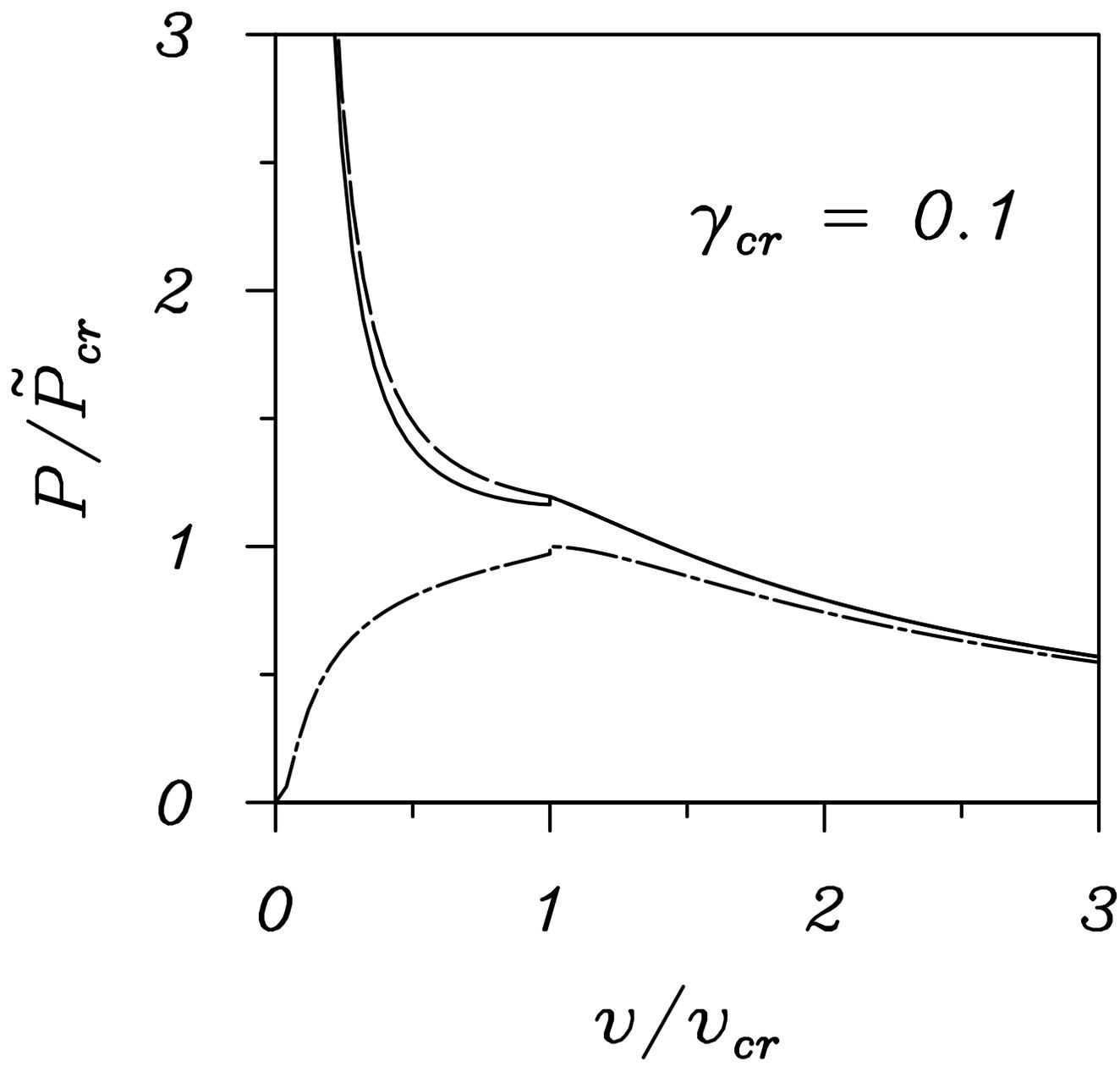
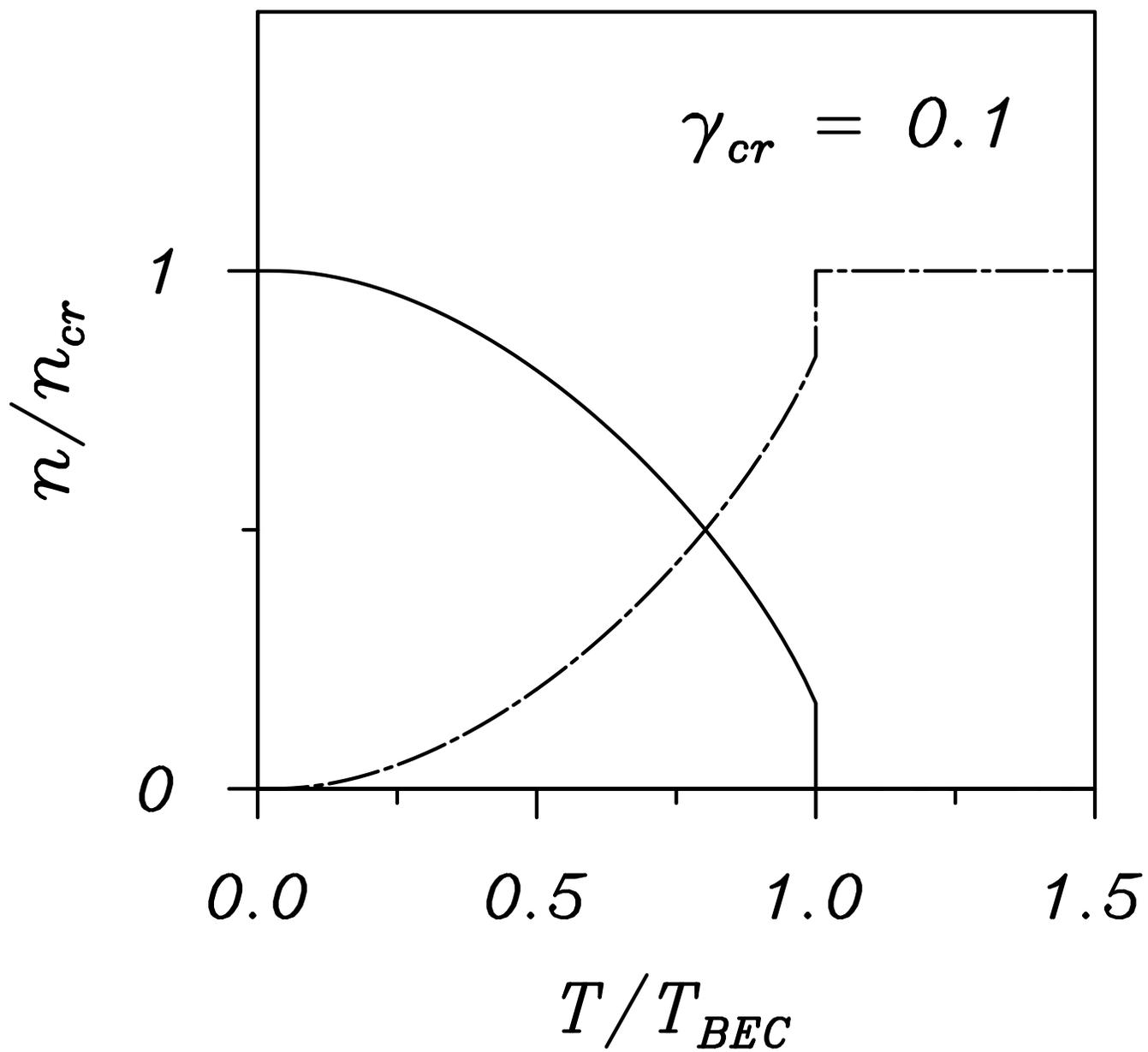
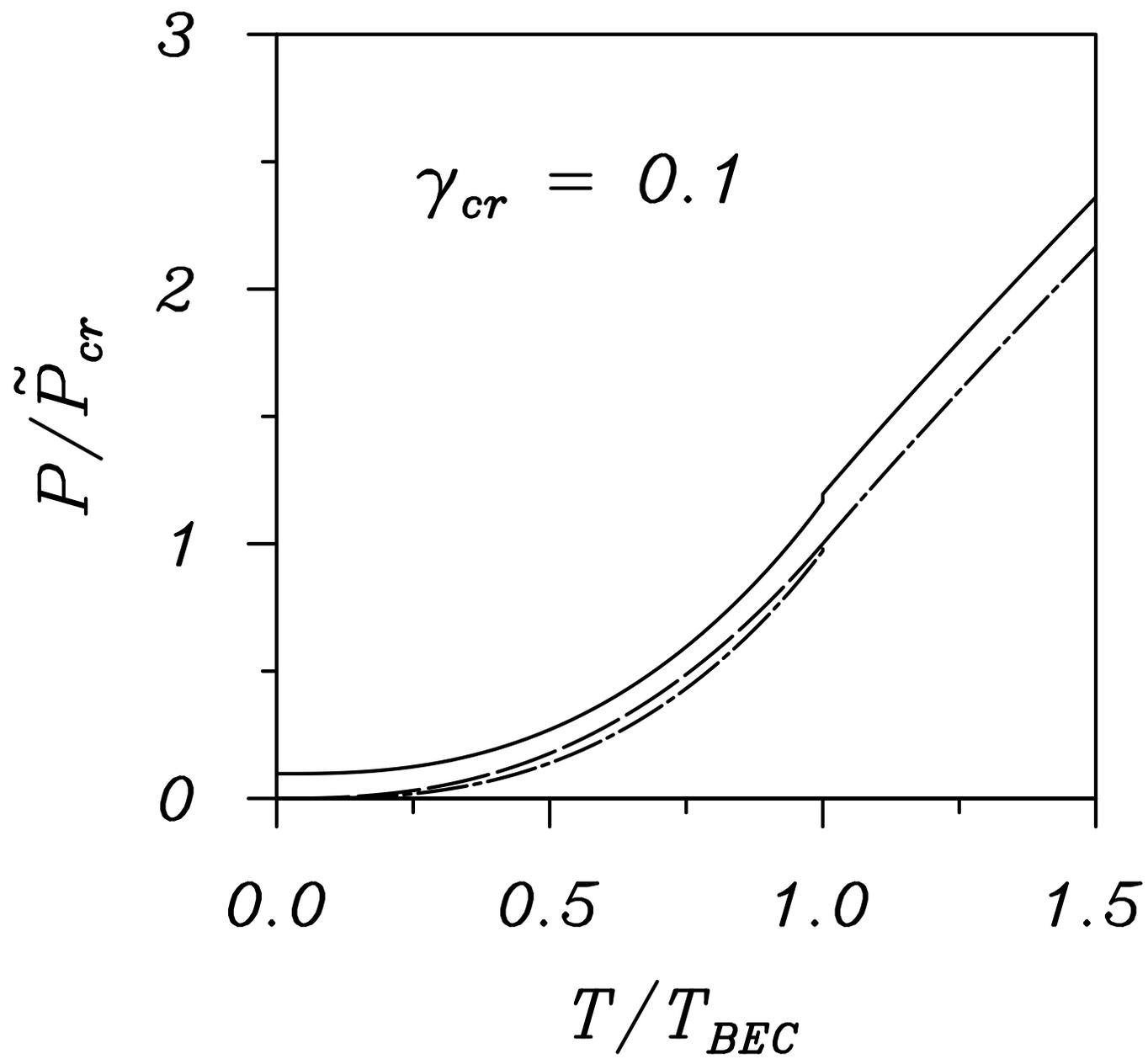
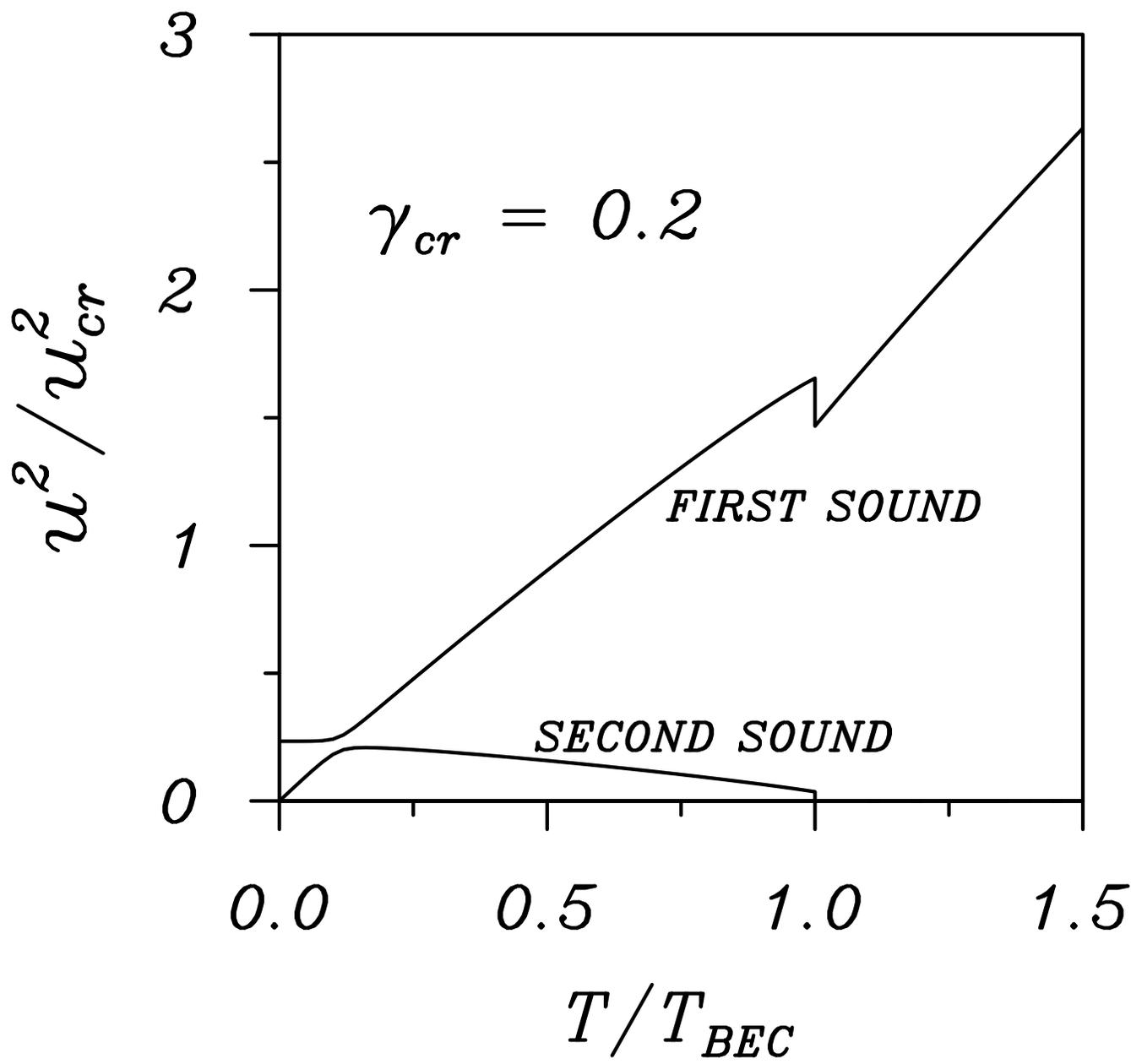
\end{document}